\documentclass[12pt]{article}
\pdfoutput=1  
\usepackage[pdftex]{graphicx}
\usepackage{amsmath,amssymb}
\usepackage{cancel}
\usepackage{appendix}
\usepackage{hyperref}
\def\zbar{\bar{z}}
\def\para{{\scriptscriptstyle ||}}

\newcommand{\tr}{\mathop{\rm Tr}}

\def\ket#1{| #1 \rangle}
\def\bra#1{\langle  #1 |}

\newcommand{\bZ}{{\mathbf{Z}}}

\newcommand{\partialbar}{\bar{\partial}}
\newcommand{\btil}{\tilde{\beta}}
\newcommand{\gtil}{\tilde{\gamma}} 
\newcommand{\braket}[2]{\langle #1|#2\rangle} 

\begin{document}
\begin{titlepage}

\begin{center}
\Large \bf \textbf{Geometry from D-branes\\ 
in Nonrelativistic String Theory}
\end{center}

\begin{center}
Alberto G\"uijosa$^{\ast}$
and Igmar C.~Rosas-L\'opez$^{\dagger}$

\vspace{0.2cm}
$^{\ast}\,$Departamento de F\'{\i}sica de Altas Energ\'{\i}as, \\
Instituto de Ciencias Nucleares, \\
Universidad Nacional Aut\'onoma de M\'exico,
\\ Apartado Postal 70-543, CdMx 04510, M\'exico\\
\vspace{0.2cm}
$^{\dagger}\,$Área de Física\\
Departamento de Preparatoria Agrícola, \\
Universidad Aut\'onoma Chapingo,
\\ km.~38.5 Carretera M\'exico-Texcoco, \\
Texcoco, Edo.Mex. 56230, M\'exico\\
 \vspace{0.2cm}
\vspace{0.2cm}
{\tt alberto@nucleares.unam.mx, 
irosasl@chapingo.mx
}
\vspace{0.2cm}
\end{center}

\begin{center}
{\bf Abstract}
\end{center}
\noindent
Nonrelativistic (NR) string theory was discovered as a framework that underlies and unifies the various noncommutative open string (NCOS) theories, which were originally envisioned as surprising exceptions to the maxim that all string theories are gravitational in nature. In that view, the fact that NCOS has a gravitational dual was believed to be directly analogous to the AdS/CFT correspondence. When NCOS theories were understood to be simply the particular classes of states of the underlying NR theory that include longitudinal D-branes, it was suggested that the duality between NCOS and the corresponding gravitational theory is not an instance of gauge/gravity-type duality, but of open-string/closed-string duality between D-branes and black branes. The present paper provides direct evidence in support of this perspective, by starting from a stack of D-branes in NR string theory and deriving the long-distance profile of the curved geometry in the corresponding black brane. 
 \vspace{0.4cm}
\smallskip
\end{titlepage}

\tableofcontents
\section{Introduction and Summary}

Following the discovery of the AdS/CFT correspondence \cite{Maldacena:1997re,Gubser:1998bc,Witten:1998qj}, it became natural to seek a dual gravity description for each non-gravitational theory with appropriate features. This is more easily attainable for theories defined through low-energy limits of string theory that lead to subsectors of states consistently decoupled from gravity.  The derivation proceeds by starting with a correspondence between two alternative descriptions that exist already at the level of the full string theory, and then applying the decoupling limit to deduce a duality between the simplified, low-energy versions of the two descriptions. This is of course the very method employed in \cite{Maldacena:1997re,Itzhaki:1998dd}, beginning from the open/closed duality between D-branes \cite{Dai:1989ua,Horava:1989ga} and black branes \cite{Horowitz:1991cd} asserted by Polchinski \cite{Polchinski:1995mt}.  

Among a plethora of instantiations of this type of reasoning, one relevant example arose from the understanding of noncommutative Yang-Mills (NCYM) theories as a limit of string theory in the presence of D-branes with a magnetic field \cite{Sheikh-Jabbari:1997qke,Sheikh-Jabbari:1998aur,Seiberg:1999vs}. The gravity dual to NCYM was promptly formulated in \cite{Hashimoto:1999ut,Maldacena:1999mh,Alishahiha:1999ci}. Follow-up work soon led to the discovery of noncommutative open string (NCOS) theories as a low-energy limit of D-branes with a constant near-critical electric field \cite{Seiberg:2000ms,Gopakumar:2000na}. Physically, such D-branes describe a D$p$ $+$ fundamental string (F1) bound state \cite{Witten:1995im,Arfaei:1997hb}. The electric field, or equivalently, the string winding within the bound state, singles out a preferred direction, which we will denote $x^1$.  NCOS theories live on the $p+1$ worldvolume dimensions of the D-branes, display space-time noncommutativity between $x^1$ and $x^0$, and were initially interpreted as having the unprecedented feature of being quantum relativistic string theories without closed strings, and consequently, without gravity. With this mindset, their gravity duals were obtained in \cite{Seiberg:2000ms,Harmark:2000wv}.  

A surprise came when \cite{Klebanov:2000pp} found that, upon compactifying $x^1$, the binding energy of the fundamental strings in the D$p$-F1 bound state is finite even after the low-energy limit, implying that, after all, there do exist closed strings in NCOS theories. To remain in the theory post-limit, these strings must  necessarily have positive winding along $x^1$ ($w>0$), so graviton ($w=0$) states appeared to be excluded, and the no-gravity feature seemed to be retained. As a result of the limit, these wound closed strings display a nonrelativistic spectrum. 

The situation was further clarified in \cite{Danielsson:2000gi, Gomis:2000bd}, which showed that the limit under consideration makes  sense for compact $x^1$ even when no D-branes and no electric (or Kalb-Ramond) field are present, in which case one is left only with the positively-wound closed strings of \cite{Klebanov:2000pp}, a setup that can be rightly called nonrelativistic closed string (NRCS) theory. But the more important conclusion is that NRCS, as well as NCOS$_{p+1}$ for all $p$, are in fact just different subsectors of a single underlying critical (10- or 26-dimensional) string theory, which has come to be known as nonrelativistic (NR) string theory.\footnote{The NR moniker \cite{Gomis:2000bd} is fitting, because all of the excitations of the theory except those of the NCOS sector are indeed nonrelativistic. An alternative but ultimately disfavored denomination, `wound string theory', was proposed in \cite{Danielsson:2000gi} to underscore the fact that all states in the theory (including those in the NCOS sector) carry positive F1 winding along the $x^1$ direction.}

The properties of NR string theory were studied in \cite{Danielsson:2000gi} with an extrinsic approach, whereby one starts with the feature of interest in the original, parent string theory and then follows the effect of the NCOS/NRCS (henceforth NR) limit. As a complement to this, the authors of \cite{Gomis:2000bd} were able to develop a very useful intrinsic approach, where the limit is taken once and for all at the level of the string worldsheet action. The resulting action provides a self-contained definition of NR string theory, where one need no longer refer to any limit. 

Soon thereafter, important subtleties of both the extrinsic and intrinsic approaches were highlighted in \cite{Danielsson:2000mu}. These included the observation that the intrinsic action of \cite{Gomis:2000bd} can give erroneous results for on-shell strings that have non-positive winding ($w\le 0$). Along with this came the realization that NR string theory does in fact contain massless string states with zero winding, including gravitons, photons and D-brane collective coordinates. On shell, unwound strings are restricted to having vanishing momentum along the spatial directions transverse to $x^1$ \cite{Danielsson:2000gi}, and are consequently negligible (measure-zero) as asymptotic states \cite{Danielsson:2000mu}. But off shell, their transverse momentum can be arbitrary, and on account of being the only massless states in the theory, they act as the mediators of all long-range interactions \cite{Gomis:2000bd,Danielsson:2000mu}. As a result of the limit, the interactions they mediate are instantaneous.
Ultimately, then, NR string theory (and therefore NCOS and NRCS) is (are) no exception to the maxim that string theory entails gravity. But it does (and they do) have the fascinating feature of yielding gravity in Newtonian guise \cite{Gomis:2000bd,Danielsson:2000mu}.
For this and other reasons, over the past two decades NR string theory has been explored and generalized in many different fruitful directions, recently reviewed in  \cite{Oling:2022fft}.

Another aspect scrutinized in \cite{Danielsson:2000mu} was the relation between NCOS and its gravity dual. It was explained there that the curved geometry of the bulk description shares with AdS the property of being effectively a box that confines massive particles but allows massless particles to reach the conformal boundary in finite time. But notably, the additional presence of a nontrivial profile of the Kalb-Ramond field $B_{01}$ leads to a very different effect for \emph{strings}: those with $w\le 0$ along $x^1$ are confined by the background, while those with $w>0$, despite being massive,  are not confined, and can in fact move out to infinity. The energy cost for this escape was shown to match the one incurred in NCOS when a $w>0$ string is freed from the D$p$-F1 bound state. In the asymptotic region of the curved background, such strings sense a flat NR geometry.  

All this was interpreted in \cite{Danielsson:2000mu} to mean that, even though at first sight the NCOS limit used to obtain the dual supergravity background \cite{Gopakumar:2000na,Harmark:2000wv} appears akin to the Maldacena near-horizon limit \cite{Maldacena:1997re}, it is actually radically different in nature.
Starting from the standard open-D-brane/black-brane (i.e., open-string/closed-string) correspondence in the parent string theory \cite{Polchinski:1995mt}, the effect of the low-energy limit on the open-string side is to yield a stack of D-branes in NR string theory, whose excitations are described by the NCOS open strings that inhabit the stack \cite{Seiberg:2000ms,Gopakumar:2000na}, but also by positively-wound closed strings that can exit the stack and move arbitrarily far away from it \cite{Klebanov:2000pp,Danielsson:2000gi,Gomis:2000bd}. In parallel with this, the effect of the limit on the closed-string side is to yield the corresponding black brane, i.e., the macroscopic field configuration sourced by the D-brane stack in question. Crucially, this configuration includes both a throat and an asymptotic region. Unwound strings on this background cannot exit the throat, so they are dual to the open strings on the NCOS D-brane stack. Positively-wound closed strings, on the other hand, can move away from the throat and into the asymptotic region, so they are dual to the wound closed strings that can depart from the D-brane stack. In short, the limit projects D-brane/black-brane duality in the parent theory down to D-brane/black-brane duality in NR string theory.

Our goal in this paper is to provide very direct evidence for this proposal, by starting with D-branes in NR string theory, described in the intrinsic formalism of \cite{Gomis:2000bd}, and verifying explicitly whether or not they give rise to the asymptotic region of the curved geometry derived in \cite{Gopakumar:2000na,Harmark:2000wv}. A successful match would prove that the background in question is nothing more and nothing less than the relevant black brane in NR string theory. The result we are after is analogous to the classic computations of scattering amplitudes and absorption cross sections \cite{Gubser:1996wt,Callan:1996tv,Garousi:1996ad,Hashimoto:1996bf,Klebanov:1997kc} that firmly established the D-brane/black-brane correspondence in relativistic string theory, and eventually led to the discovery of the AdS/CFT correspondence \cite{Maldacena:1997re,Gubser:1998bc,Witten:1998qj} as a low-energy limit thereof.\footnote{For discussion of open/closed duality in the intermediate-energy regime where the closed side admits a supergravity description but the D-brane stack is not yet decoupled from its surrounding spacetime, see \cite{Danielsson:2000ze,Amador:2002is,Amador:2003ju}.}  Technically though, it will be more efficient for us to employ the boundary state description of D-branes \cite{Cremmer:1972gak,Clavelli:1973uk,Callan:1987px,Green:1994ix} 
to extract from it the long-range profile of the closed string fields they source, in parallel with what was done in \cite{DiVecchia:1997vef,DiVecchia:1999uf,DiVecchia:1999mal,DiVecchia:2003ne} within standard string theory.  

Our paper is structured as follows. Section \ref{preliminariessec} gives a brief review of the Gomis-Ooguri \cite{Gomis:2000bd} formulation of NR string theory. In Section~\ref{longitudinalsec} we then determine the boundary state associated with, and long-distance geometry produced by,  D-branes extended along $x^1$, which we will refer to as longitudinal. Successful agreement is found with the corresponding black brane background. The calculation is then repeated in Section \ref{transversesec} for transverse D-branes (i.e., those localized along $x^1$), which were first studied in \cite{Danielsson:2000mu}. The associated black brane background is derived here  for the first time, at the beginning of Section \ref{bkgdtranssubsec}. It has the peculiarity of being itself delta-function localized along $x^1$, but having a nontrivial profile in the remaining directions not spanned by the brane.
Even with this feature included, a perfect match is again achieved with the long-distance fields sourced by transverse D-branes. 

All in all, then, our results give further credence to the realization that the supergravity backgrounds under consideration ought to be regarded as the black branes of NR string theory.\footnote{The analogous statement is then implied as well for all the various NR (also described as Galilean or, originally, `wrapped') brane theories that are related to NR string theory through a web of dualities \cite{Danielsson:2000gi,Gomis:2000bd}. In particular, NCOS$_{3+1}$ is S-dual to NCYM$_{3+1}$, which is in turn T-dual to NCYM$_{p+1}$ \cite{Gopakumar:2000na} One of the implications then is that the latter theories are not really ($p+1$)-dimensional, because D1-branes can become unbound from their host D$p$ stack at finite energy cost, and then move away to infinity \cite{Danielsson:2000gi}.}  
It might seem puzzling to some readers that these backgrounds are not of string Newton-Cartan form, and are consequently not solutions to the curved-space equations of motion that have been extensively studied in recent years as interesting generalizations of flat-space NR string theory \cite{Harmark:2017rpg,Bergshoeff:2018yvt,Harmark:2018cdl,Gomis:2019zyu,Harmark:2019upf,Gallegos:2020egk,Bergshoeff:2021bmc,Hartong:2021ekg,Bidussi:2021ujm} (for more references, see \cite{Oling:2022fft}). We will elaborate on this issue in a separate publication \cite{Avila:2023}.

\section{NR string theory} \label{preliminariessec}
Nonrelativistic string theory on a flat $d$-dimensional spacetime can be intrinsically defined by the Gomis-Ooguri action \cite{Gomis:2000bd}, which in conformal gauge takes the form
\begin{equation}
S_{\text{GO}}=\int \frac{d^2z}{2\pi}\left(  \beta\partialbar\gamma +\btil\partial\gtil +\frac{\mu}{2} \partial\gamma\partialbar\gtil +\frac{1}{L_s^2}\partial X^j\partialbar X_j\right)~,\label{actiongo}
\end{equation}
where $L_s$ is the string length, $X^j$ denotes the string embedding fields along the transverse directions $x^j$ with $j=2,\cdots, d-1$,  and 
\begin{equation}
L_s\gamma\equiv X^0+X^1~, 
\quad 
L_s\gtil\equiv-X^0+X^1
\end{equation}
are a lightcone-coordinate repackaging of the embedding fields along the longitudinal spacetime directions $x^a$, with $a=0,1$. To allow for closed string states with finite energy, we take the $x^1$ direction to be compactified on a circle of radius $R$ \cite{Klebanov:2000pp,Danielsson:2000gi,Gomis:2000bd}. 
The $\beta$, $\tilde\beta$ fields, conjugate to $\gamma,\tilde\gamma$, are auxiliary variables of the theory, which play the role of Lagrange multipliers \cite{Gomis:2000bd}. The parameter $\mu$ (arising from a shift in the Kalb-Ramond field in the parent string theory) is a constant whose value is completely inconsequential for closed strings or open strings attached to transverse D-branes (as in Section~\ref{transversesec}) \cite{Danielsson:2000gi,Danielsson:2000mu}. Setting $\mu=0$ reduces the longitudinal part of the action to a standard commuting ghost system, with conformal weights $h_{\gamma}=0$, $h_{\beta}=1$. On the other hand, when considering open strings attached to longitudinal D-branes (as in Section~\ref{longitudinalsec}), $\mu$ takes on a definite non-zero value, given by (\ref{mulongitudinal}) below  \cite{Danielsson:2000gi,Danielsson:2000mu}.

The equations of motion ensuing from varying the action (\ref{actiongo}) are
\begin{equation}
\partialbar\gamma =\partial\gtil=\partialbar\beta=\partial\btil= \partial\partialbar X^j=0~.
\end{equation}

Considering the usual radial quantization on the complex plane, for a closed string one enforces the natural periodicity conditions
\begin{align}
\gamma(e^{2\pi i}z)-\gamma(z)&=\frac{2\pi wR}{L_s}~. 
\nonumber \\
\tilde\gamma(e^{-2 \pi i}\bar z)-\tilde\gamma(\bar z)&=\frac{2\pi wR}{L_s}~.
\nonumber\\
X^i(e^{2 \pi i}z,e^{-2 \pi i}\bar z)&=X^i(z,\bar z)~,\\
\beta(e^{2 \pi i}z)&=\beta(z)~,
\nonumber\\
\tilde\beta(e^{-2 \pi i}\bar z)&=\tilde\beta(z)~,
\nonumber
\end{align}
where $w>0$ 
is the string winding along the compact $x^1$ direction.\footnote{As explained in \cite{Danielsson:2000mu}, the derivation of the action (\ref{actiongo}) in \cite{Gomis:2000bd} implicitly assumes positive string winding, and therefore leads to physically incorrect results for on-shell closed strings with $w\le 0$.}

The mode expansions for the transverse embedding fields  $X^i$ are the familiar ones.  For the longitudinal embedding fields, one has
\begin{align}
\beta(z)&=\sum_{n=-\infty}^\infty\beta_nz^{-n-1}~,
\quad \btil(\zbar)=\sum_{n=-\infty}^\infty\btil_n{\zbar}_n^{-n-1}~,
\nonumber\\
\gamma(z,\zbar)&=+i\frac{w R}{L_s}\log z+\sum_{n=-\infty}^\infty \gamma_n z^{-n}~, \\
\gtil(z,\zbar)&=-i\frac{w R}{L_s}\log\zbar+\sum_{n=-\infty}^\infty\gtil_n{\zbar}^{-n}~.
\nonumber
\end{align}
The non-zero conmutators are
\begin{equation}
[\gamma_n,\beta_m]=\delta_{n+m}~, \quad [\gtil_n,\btil_m]=\delta_{n+m}~.
\end{equation}
The contribution from the longitudinal directions to the energy-momentum tensor reads
\begin{equation}
T^{{\mbox{\tiny long}}}(z)=-:\beta\partial\gamma:~,\quad \tilde{T}^{{\mbox{\tiny long}}}(\zbar)=-:\btil\partialbar\gtil:~.
\end{equation}
Using this, we can obtain in particular the full Virasoro zero modes
\begin{equation}
\begin{split}
L_0 &=   -i\frac{w R}{L_s} \beta_0+\sum_{n >0} n (\beta_{-n}\gamma_n -\gamma_{-n}\beta_n)+
\frac{L_s^2p_{{\mbox{\tiny trans}}}^2}{4}+\sum_{n > 0} \alpha_{-n}^i\alpha_n^i~, \\
\tilde{L}_0 &= ~~  i \frac{w R}{L_s} \tilde{\beta}_0+\sum_{n>0} n (\btil_{-n}\gtil_n-\gtil_{-n}\btil_n)+\frac{L_s^2p_{{\mbox{\tiny trans}}}^2}{4}+\sum_{n>0} {\tilde{\alpha}}_{-n}^i{\tilde{\alpha}}_n^i ~,
\end{split}
\label{ellzerotot}
\end{equation}
where $p_{{\mbox{\tiny trans}}}$ is the momentum along the transverse directions, and $\beta_0$, $\tilde\beta_0$ are related to the momenta $p_0$, $p_1$ conjugate to $x^0$, $x^1$ through
\begin{align}
L_s p_+&\equiv \frac{L_s}{2}(+p_0+p_1)=i\beta_0+\frac{\mu}{2}\left(\frac{w R}{L_s}\right)~,\\
L_s p_- &\equiv  \frac{L_s}{2}(-p_0+p_1)=i\btil_0-\frac{\mu}{2}\left(\frac{wR}{L_s}\right)~.
\end{align}
The periodicity of $x^1$ implies that $p_1=\frac{n}{R}$.

For an open string, the variational principle demands that boundary conditions be chosen such that, at $z=\bar{z}$,
\begin{equation}
\begin{split}
\delta\gamma(\beta-\frac{\mu}{2}\partialbar\gtil)+\delta\gtil(-\btil+\frac{\mu}{2}\partial\gamma)&=0~, 
\\
\delta X^i(\partial -\bar\partial)X^i~. 
\end{split}
\label{fronteraWSTabierta}
\end{equation}
As in standard string theory, these requirements can be satisfied with either Dirichlet or Neumann boundary conditions, which are appropriate for open strings which are excitations of D-branes with different dimensionalities and orientations. 

Given the salient distinction made in NR string theory between longitudinal and transverse directions, when addressing (\ref{fronteraWSTabierta}) the crucial issue is whether or not the D-brane is extended along the $x^1$ direction. We describe the two possibilities separately in Sections~\ref{longitudinalsec} and~\ref{transversesec}, which contain the main calculations of the present paper. 

\section{Longitudinal D-branes}
\label{longitudinalsec}

In this section we will work out the description of a longitudinal D-brane stack in terms of the boundary state formalism \cite{Cremmer:1972gak,Clavelli:1973uk,Callan:1987px,Green:1994ix}. This originates from considering an annulus amplitude and performing a conformal transformation that exchanges the roles of space $\sigma$ and time $\tau$ on the string worldsheet, so that the requisite open string boundary conditions are reinterpreted as \emph{initial} conditions for a \emph{closed} string. One thus seeks to construct a state $\ket{B}$ in the closed string Hilbert space that complies with the given initial conditions. Such `boundary state' is of coherent type, and automatically encodes the amplitude for the D-brane to emit a closed string with any given oscillator level $N$. In particular, as we will show, the long-distance profile of the metric and Kalb-Ramond fields sourced by the D-brane can be extracted simply by computing the relevant one-point functions at massless level. 

To derive the boundary state and the asymptotic gravity fields that arise from it, in this section and the next we will follow a procedure that closely parallels the one developed by Di Vecchia \emph{et al.} in the relativistic string setting \cite{DiVecchia:1997vef,DiVecchia:1999uf,DiVecchia:1999mal,DiVecchia:2003ne}.

As is well-known, the massless closed string sector of the relativistic bosonic string is also present in the superstring theories, even though the oscillator makeup of these states is different in each case. Upon taking the limit to reach the corresponding NR string theories, the same commonality is present.\footnote{The fermionic analog of the Gomis-Ooguri action (\ref{actiongo}) was first derived in \cite{Danielsson:2000mu}. Quite naturally, its longitudinal piece is simply the anticommuting counterpart of the $\beta$-$\gamma$ system, i.e., a $b$-$c$ ghost system with weights $h_c=h_b=1/2$. Other work on supersymmetric NR string theory can be found in \cite{Gomis:2004pw,Gomis:2005pg,Kim:2007hb,Kim:2007pc,Park:2016sbw,Blair:2019qwi}.} For definiteness, we will carry out our computations in the notation appropriate for the NR bosonic string, keeping the spacetime dimension $d$ arbitrary, for ease of transitioning to the superstring theories. This shortcut is justified by the agreement found in \cite{DiVecchia:1997vef,DiVecchia:1999uf,DiVecchia:1999mal,DiVecchia:2003ne} between the relativistic bosonic string calculation extrapolated to $d=10$ and the corresponding black brane in superstring theory. Such agreement is naturally expected to persist upon taking the NR limit of both descriptions.  

\subsection{Boundary conditions}
\label{bclongsubsec}
 
By definition, for open strings attached to a longitudinal D-brane stack we have $\delta\gamma,\delta\tilde\gamma\neq 0$ in
 (\ref{fronteraWSTabierta}), so the variational principle dictates the boundary conditions
\begin{equation}
\left(\beta-  \frac{\mu}{2}\partialbar\gtil  \right)\Bigg{|}_{z=\zbar}=0,\quad\quad \left(  \tilde\beta -\frac{\mu}{2}\partial\gamma   \right)\Bigg{|}_{z=\zbar}=0~, \label{longbcond}
\end{equation}
\begin{align}
(\partial-\bar\partial)X^i|_{z=\bar z}&=0 ~  \quad\quad  i =2,\dots ,p \ , \\
X^i |_{z=\bar z} &=y^i\quad\quad i=p+1,\dots,d-1~,
\end{align}
where $y^i$ labels the position of the stack. Importantly, for this setup the parameter $\mu$ in (\ref{actiongo}) and (\ref{longbcond}) is directly physical: it is determined by the total D$p$ and F1 charges of the stack, and takes the specific value \cite{Danielsson:2000gi,Danielsson:2000mu}
\begin{equation}
\mu=\frac{K^2}{2\nu^2G_s^2}~,
\label{mulongitudinal}
\end{equation}
with $G_s$ the string coupling, $K$ the number of coincident D-branes and $\nu$ the density of fundamental strings bound to them (related to the number of units of electric flux).

The mode expansions obtained from the boundary conditions (\ref{longbcond})  are
\begin{align}
\beta(z)=\sum_{n=-\infty}^\infty\beta_n z^{-n-1}&\Rightarrow\gtil(\zbar)=\frac{2}{\mu}\beta_0\ln\zbar +\gtil_0-\frac{2}{\mu}\sum_{n\neq 0}\frac{\beta_n}{n}\zbar^{-n} \\
\btil(\zbar)=\sum_{n=-\infty}^\infty\btil_n {\zbar}^{-n-1}&\Rightarrow\gamma(z)=\frac{2}{\mu}\btil_0\ln\zbar +\gamma_0-\frac{2}{\mu}\sum_{n\neq 0}\frac{\btil_n}{n}z^{-n}~.\nonumber
\end{align}
The longitudinal part of the energy-momentum tensor and the Virasoro operators take the form
\begin{align}
T^{{\mbox{\tiny long}}}(z)&=\frac{2}{\mu}:\beta(z)\btil(\zbar): ~,\\
L_n^{{\mbox{\tiny long}}}&=\frac{2}{\mu}\sum_l:\beta_l\btil_{n-l}:~.
\end{align}

Here, the non-zero conmutators for the open string modes are
\begin{equation}
[\gamma_0 ,\beta_0]=[\gtil_0,\btil_0]=1,\quad [\btil_n,\beta_m]=\frac{\mu}{2}n\delta_{n+m}~.
\end{equation}


\subsection{Boundary state}
\label{bdrystatelongsubsec}


The direct interpretation of an annulus amplitude is as a loop of a string with endpoints at $\sigma=0,\pi$.
After a conformal transformation that interchanges $\sigma\leftrightarrow\tau$, the amplitude is reinterpreted as a cylinder that computes tree-level propagation of a closed string
\cite{DiVecchia:1997vef,DiVecchia:1999uf,DiVecchia:1999mal,DiVecchia:2003ne}. The open string boundary condition applied at $\sigma=0$ is mapped to an initial condition enforced on the closed string state $\ket{B}$ at $\tau=0$. In the usual complex coordinates, the latter location corresponds to the unit circle, i.e., $z=\bar z^{-1}$.

{}Under the mapping, the specific boundary conditions (\ref{longbcond}) for a longitudinal D-brane located at $x^i=y^i$ ($i=p+1,\ldots,d-1$) are transmuted into the following initial conditions for the boundary state:
\begin{align}
\left(\frac{\mu}{2}z\partial\gamma+\zbar\btil \right)\Big{|}_{z={\zbar}^{-1}}\ket{B^{\,y}_{\mbox{\tiny long}}}&=0 ~,
\nonumber\\
\left( \frac{\mu}{2}\zbar\partialbar\gtil+z\beta \right)\Big{|}_{z={\zbar}^{-1}}\ket{B^{\,y}_{\mbox{\tiny long}}}&=0 ~,\\
(z\partial+\bar z\bar\partial) X^i|_{z=\bar z^{-1}}\ket{B^{\,y}_{\mbox{\tiny long}}}&= 0 ~\qquad i=2,\dots,p \ , 
\nonumber\\
\left(X^i-y^i\right) |_{z=\bar z^{-1}}\ket{B^{\,y}_{\mbox{\tiny long}}} &= 0 \qquad i=p+1,\dots , d-1 \ .
\nonumber
\end{align}
We have introduced an explicit label in the ket to indicate the location of the D-brane. Rewriting the boundary conditions in terms of the  modes, we get
\begin{align}
\left(  \frac{\mu n}{2}\gamma_n-\btil_{-n}  \right)\ket{B^{\,y}_{\mbox{\tiny long}}}&=0\quad  \quad \forall~n\neq 0~,
\nonumber\\
\left(  \frac{\mu n}{2}\gtil_n-\beta_{-n}  \right)\ket{B^{\,y}_{\mbox{\tiny long}}}&=0 \quad  \quad \forall~n\neq 0~, 
\nonumber\\
\left(  \frac{i\mu w R}{2L_s}+\btil_0     \right) \ket{B^{\,y}_{\mbox{\tiny long}}} &=0 ~,
\nonumber\\
\left(  \frac{i\mu w R}{2L_s}-\beta_0     \right) \ket{B^{\,y}_{\mbox{\tiny long}}} &=0~, \\
(\alpha_n^i+{S^i}_j\tilde\alpha_{-n}^j)\ket{B^{\,y}_{\mbox{\tiny long}}}  &= 0~~~~~\forall~n \neq 0
~, 
\nonumber\\
{\hat{p}}^i \ket{B^{\,y}_{\mbox{\tiny long}}}   &= 0  \ \ \ \ i=2,\dots ,p~,
\nonumber\\
({\hat{x}}^i-y^i)\ket{B^{\,y}_{\mbox{\tiny long}}}  &= 0~ \ \ \ i=p+1,\dots, d-1~,
\nonumber
\end{align}
where we have introduced the $(d-2)\times(d-2)$ diagonal matrix
\begin{equation}
S^{ij}=\left\{ \begin{array}{ll}
~~\delta^{ij}  &  \text{if } \ \ i,j =2,\dots , p \\
-\delta^{ij} & \text{if } \ \ i,j =p+1,\dots ,d\qquad~. \\
~~0 &  \text{otherwise}
\end{array}\right.\label{Sijlong}
\end{equation}

In the presence of a longitudinal D-brane, closed string winding is by itself not conserved: a closed string with winding $w$ may break open, have its endpoints move around the $x^1$ circle some integer number $m$ of times, and then reconnect as a closed string with winding $w'=w-m$. This is compensated by the fact that in the course of this process the electric charge on the open string endpoints changes the number of units of electric flux  on the D-brane from $e$ to $e'=e+m$, which is physically equivalent to binding $m$ fundamental strings to the brane. So,  altogether, the \emph{total} F1 winding number $W\equiv w+e$ is conserved. For our purposes, the important feature is that a longitudinal D-brane can emit closed strings of \emph{any} winding. The boundary state that describes this must therefore include a sum over all possible windings--- unrestricted by the $w>0$ condition, which applies only to on-shell closed string states. 

The state that fulfills all the given conditions can be determined to be
\begin{equation}
\ket{B^{\,y}_{\mbox{\tiny long}}} = N_p\sum_{w =-\infty}^\infty\delta^{(d-p-1)}({\hat{x}}^i-y^i)\left(\prod_{n=1}^\infty e^{-\frac{1}{n}\alpha_{-n}\cdot S\cdot{\tilde{\alpha}}_{-n}+\frac{2}{n\mu}\beta_{-n}{\btil}_{-n} +\frac{\mu n}{2}\gamma_{-n}{\gtil}_{-n}}\right)\ket{0,w, p=0},  \label{StateLong}
\end{equation}
with $N_p$ a normalization constant, which naturally depends on the D$p$ and $F1$ charges of the specific D-brane stack under consideration, i.e., on $K$ and $\nu$.

\subsection{Normalization of boundary state}
\label{normalizationsubsec}
In the expression for the boundary state there is an undetermined constant $N_p$.  In order  to compare the result  coming from  the calculation in terms of the boundary state formalism with the results in the black brane background, it is essential to establish  the normalization.

For this purpose, similar to the classic calculation in \cite{Polchinski:1995mt}, we will compute  the interaction between two parallel D$p$-branes located respectively at $x^i=0$ and $x^i=y^i$, both in the open and closed string channels. By comparing both results, we will obtain the normalization constant $N_p$.

In the closed string channel, the cylinder amplitude describes the interaction between two parallel D-branes, given by
\begin{equation}
\mathcal{A}_{\mbox{\tiny closed}}=\langle B^{\,0}_{\mbox{\tiny long}}|D|B^{\,y}_{\mbox{\tiny long}}\rangle ~,
\end{equation}
with $D$ the free string propagator
\begin{equation}
D=\frac{L_{s}^{2}}{8\pi}\int\limits_{|z|\leq 1}\frac{d^2z}{{|z|}^2} z^{L_0-1}{\zbar}^{{\tilde{L}}_0-1}=\frac{L_{s}^{2}}{8\pi}\int\limits_{|z|\leq 1}\frac{d^2z}{{|z|}^4} z^{L_0}{\zbar}^{{\tilde{L}}_0}~,
\end{equation}
where $L_0$, $\tilde{L}_0$ are the closed-string Virasoro operators (\ref{ellzerotot}).
%
Propagating the boundary state, we get
\begin{align}
D\ket{B^{\,y}_{\mbox{\tiny long}}} &= \frac{ L_s^2}{8\pi} \int_{\mid z\mid \leq 1}\frac{d^2z}{{|z|}^4}z^{L_0}{\bar{z}}^{\tilde L_0}\ket{B_{\mbox{\tiny long}}} \nonumber \\
&=  \frac{N_p L_s^2}{8\pi} \int_{\mid z\mid \leq 1}\frac{d^2z}{{|z|}^4}z^{L_0}{\bar{z}}^{\tilde L_0}\sum_{w=-\infty}^\infty \delta^{(d-p-1)}(\hat{x}^i-y^i)
\nonumber  \\
& \      \left(\prod_{n=1}e^{-\frac{1}{n}\alpha_{-n}S\cdot\tilde{\alpha}_{-n}+\frac{2}{n\mu }\beta_{-n}\tilde{\beta}_{-n}+\frac{n\mu}{2}\gamma_{-n}\tilde{\gamma}_{-n}}\right)\ket{0, w ,p=0} 
  \\
&= \frac{N_p L_s^2}{8\pi} \int\limits_{|z| \leq 1}\frac{d^2z}{{|z|}^4}\int\frac{d^{d-p-1}k_\perp}{(2\pi )^{d-p-1}}|z|^{\frac{L_s^2}{2}k_{\perp}^2}e^{-ik_\perp\cdot y}\sum_{w=-\infty}^\infty  |z|^{\frac{\mu w^2 R^2}{L_s^2}} 
\nonumber\\
& \   \left(\prod_{n=1}e^{(-\frac{1}{n}\alpha_{-n}S\cdot\tilde{\alpha}_{-n}+\frac{2}{n\mu}\beta_{-n}\tilde{\beta}_{-n}+\frac{n\mu}{2}\gamma_{-n}\tilde{\gamma}_{-n})|z|^{2n}}\right)
\ket{0, w ,p_\para=0,p_\perp = k_\perp}
\nonumber  ~,
\end{align}
where $p_\para$ and $p_\perp$ denote respectively the directions parallel and perpendicular to the D$p$-brane stack.\footnote{It is important not to confuse these parallel ($\mu=0,1,\ldots,p$) and perpendicular ($\mu=p+1,\ldots,d-1$) directions, whose difference makes reference to the longitudinal D-branes, with the longitudinal ($\mu=0,1$) and transverse ($\mu=2,\ldots,d-1)$ directions whose distinction is intrinsic to NR string theory.} 

Projecting this on the boundary state bra at the origin, we finally obtain the amplitude in the closed string channel,
\begin{align}
\mathcal{A}_{\mbox{\tiny closed}}&=
%
\frac{N_p^2L_s^2}{8\pi} \sum_{w^\prime =-\infty}^\infty\bra{0,w^\prime, p=0}\delta^{(d-p-1)}({\hat{x}}^i)\left(\prod_{n=1}^\infty e^{-\frac{1}{n}\alpha_{n}S\cdot{\tilde{\alpha}}_{n}+\frac{2}{n\mu}\beta_{n}{\btil}_{n} +\frac{n\mu}{2}\gamma_{n}{\gtil}_{n}}\right)  \nonumber \\
&~~~~~\int\limits_{|z|\leq 1}\frac{d^2z}{{|z|}^2} z^{L_0-1}{\zbar}^{{\tilde{L}}_0-1}\sum_{w =-\infty}^\infty\delta^{(d-p-1)}({\hat{x}}^i-y^i)\Big(\prod_{n=1}^\infty e^{-\frac{1}{n}\alpha_{-n}S\cdot{\tilde{\alpha}}_{-n}}
\nonumber\\
&~~\qquad\qquad\qquad \qquad\qquad\qquad \qquad e^{\frac{2}{n\mu}\beta_{-n}{\btil}_{-n} +\frac{n\mu}{2}\gamma_{-n}{\gtil}_{-n}}\Big)\ket{0,w, p=0}
\label{closedamplitudelong}\\
&= \sum_{w=-\infty}^{\infty}\frac{N_p^2 L_s^2}{8\pi}\int\limits_{|z|\leq 1}\frac{d^2z}{|z|^4}(2\pi)^{p+1}R~\delta^{(p)}(0)\int\frac{d^{d-p-1}k_\perp}{(2\pi)^{d-p-1}}|z|^{\frac{L_s^2}{2}k_\perp^2} \nonumber \\
& \qquad\qquad\qquad\qquad\qquad\qquad\qquad\qquad e^{ik_\perp\cdot y}|z|^{\mu\frac{w^2R^2}{L_s^2}}\prod_{n=1}^\infty\left(\frac{1}{1-|z|^{2n}}\right)^{d-2}~,\nonumber
\end{align}
where we have used the normalization
\begin{equation}
\braket{k}{k^\prime}=2\pi\delta(k-k^\prime)~.
\end{equation}
Up to now, we had only shown explicitly the contribution of the string embedding fields (both longitudinal and transverse). That calculation by itself would yield a final exponent in (\ref{closedamplitudelong}) equal to $d$. In writing $d-2$ instead, we are acknowledging the additional contribution of the reparametrization ghosts $b,c$, which as usual leads us to the result that reflects the correct count of degrees of freedom \cite{DiVecchia:1997vef,DiVecchia:1999uf}.

Changing variables $|z|=e^{-\pi t}$, $d^2 z=-2\pi e^{-2\pi t}dtd\phi$ and performing the Gaussian integral, we get
\begin{align}
\mathcal{A}_{\mbox{\tiny closed}}&= 2\sum_w \frac{N^2_pL_s^2}{8}\int_0^\infty dt~(2\pi)(2\pi R)V_p\, e^{-\pi\mu\big(\frac{w R}{L_s}\big)^2t}e^{-\frac{y^2}{2\pi L_s^2 t}}
\label{closedamplitudelong2}\\
&\qquad\qquad\qquad\qquad  \left(\frac{2}{L_s^2 t}\right)^{\frac{d-p-1}{2}}(2\pi)^{1-p-d}\left[f_1(e^{-\pi t})\right]^{2-d}(e^{2\pi t})^{\frac{26-d}{24}}~,
\nonumber 
\end{align}
with
\begin{equation}
(2\pi)^d\delta^{(d)}(0)\equiv V_d~.
\nonumber
\end{equation}
When writing (\ref{closedamplitudelong2}), we have again followed \cite{DiVecchia:1997vef,DiVecchia:1999uf}
in defining the function
\begin{equation}
f_1(q)\equiv q^{\frac{1}{12}}\prod_{n=1}^{\infty}\left(1-q^{2n}\right)~,
\end{equation}
whose modular transformation  property (\ref{modular}) will be of use momentarily. 
This definition allowed us to translate 
\begin{equation}
 e^{2\pi t}\prod_{n=1}^\infty\left(\frac{1}{1-e^{-2\pi nt}}\right)^{d-2}=[f_1(e^{-\pi t})]^{2-d}(e^{2\pi t})^{\frac{26-d}{24}}~.
 \label{26}
\end{equation}

In order to be able to compare the closed string channel result (\ref{closedamplitudelong2}) with the one for the open string channel that will be determined below, we preventively perform yet another change of variables. Using Poisson's identity
\begin{equation}
\sum_{n=-\infty}^\infty e^{-\pi n^2 A +2n\pi As}=\frac{1}{\sqrt A}e^{\pi As^2}\sum_{m=-\infty}^\infty e^{-\pi A^{-1}m^2-2 \pi i ms}\label{Poisson}
\end{equation}
with $n=w$, $s=0$, $A=\frac{\mu R^2}{L_s^2}$, and changing the integration variable in (\ref{closedamplitudelong}) according to $t\rightarrow t=\frac{1}{\tau}$, with the help of the modular transformation property \cite{DiVecchia:1997vef,DiVecchia:1999uf} 
\begin{equation}
f_1(e^{-\pi/t}) =\sqrt{t}\,f_1(e^{-\pi t})
\label{modular}
\end{equation}
we finally obtain
\begin{align}
\mathcal{A}_{\mbox{\tiny closed}}&= N_p^2 V_p L_s^{-d+p+4}\mu^{-\frac{1}{2}}2^{\frac{1+p-d}{2}}\pi^{3+p-d}\label{FLongitudinalClosedChannel}\\
&  ~~~~\int_0^\infty d\tau \ \tau^{-\frac{p+2}{2}}e^{-\frac{y^2\tau}{2\pi L_s^2}}\sum_{m=-\infty}^{\infty} e^{-\frac{\pi}{\mu}\left(\frac{m L_s}{R}\right)^2\tau}\left[f_1(e^{-\pi\tau})\right]^{2-d}(e^{\frac{2\pi}{\tau} })^{\frac{26-d}{24}}~.\nonumber
\end{align}

In the open string channel, we ought to calculate the analogue of a vacuum loop in quantum field theory. We will need the open string Virasoro operator
\begin{align}
L_0&=\frac{2}{\mu}L_s^2p_+p_-+\frac{2}{\mu}\sum_{n=1}^\infty (\beta_n\btil_{-n}+\btil_n\beta_{-n})
\nonumber\\
&~~~~~~~~~~~~+L_s^2k^2_\perp +\frac{y^2}{(2\pi)^2L_s^2}+\sum_{n=1}^\infty\alpha^i_{-n}\cdot\alpha^i_n~, \\
p_+&\equiv\frac{1}{2}(+p_0+p_1)~,
\nonumber\\
p_-&\equiv\frac{1}{2}(-p_0+p_1)~.
\nonumber
\end{align}

The amplitude for the open string vacuum loop is given by
\begin{align}
\mathcal{A}_{\mbox{\tiny open}}&=-(2K^2)\frac{1}{2}\tr\log[L_0-1]=\int_0^\infty\frac{d\tau}{2\tau}\tr\left[e^{-2\pi (L_0-1)\tau}\right] 
\label{traza1}\\
&=(2K^2)\int_0^\infty\frac{d\tau}{2\tau}\int_0^\infty\frac{d^{p-1}k}{(2\pi)^{p-1}}\int\frac{dp_0}{2\pi}\sum_{n=-\infty}^{\infty}\frac{1}{2\pi R}\bra{k,p_1=\frac{n}{R},p_0}e^{2\pi\tau}e^{\frac{\pi}{\mu}p_0^2L_s^2\tau}
\nonumber \\
& \times e^{-\frac{\pi n^2L_s^2}{\mu R^2}\tau}e^{-\frac{y^2 \tau}{2\pi L_s^2}}e^{-2\pi L_s^2k^2\tau}\ket{k,p_1=\frac{n}{R},p_0}\tr\left(\prod_{n=1}^\infty e^{-2\pi\tau\left(\frac{2}{\mu}\beta_{-n}\btil_n+\frac{2}{\mu}\btil_{-n}\beta_n+\alpha_{-n}^i\alpha_n^i\right)}\right).
\nonumber
\end{align}
The factor of $2K^2$ in front is the number of distinct open strings that can extend between the $K$ D-branes at $x^i=0$ and the $K$ D-branes at $x^i=y^i$, considering the two possible orientations for the string.

The oscillator trace yields
\begin{equation}
\tr\left(\prod_{n=1}^\infty e^{-2\pi\tau\left(\frac{2}{\mu}\beta_{-n}\btil_n+\frac{2}{\mu}\btil_{-n}\beta_n+\alpha_{-n}^i\alpha_n^i\right)}\right)=\prod_{n=1}^\infty\left(\frac{1}{1-e^{-2\pi\tau n}}\right)^{d-2}~,
\end{equation}
again effecting the change $d\to d-2$ in the final exponent to account for the contribution of the ghosts \cite{DiVecchia:1997vef,DiVecchia:1999uf}.
Using this in (\ref{traza1}) translated into powers of $f_1$ as in (\ref{26}), and evaluating the Gaussian integrals (which implement the trace for the zero modes), we finally arrive at
\begin{equation}
\mathcal{A}_{\mbox{\tiny open}}=\frac{K^2 V_p}{(2\pi)^p}\frac{\mu^{1/2}}{L_s^{p}2^{\frac{p-1}{2}}}\int_0^\infty d\tau ~\tau^{-\frac{p+2}{2}}e^{-\frac{y^2 \tau}{2\pi L_s^2}}\sum_{n=-\infty}^{\infty} e^{-\frac{\pi}{\mu}\left(\frac{nL_s}{R}\right)^2\tau}\left[f_1(e^{-\pi\tau})\right]^{2-d}(e^{\frac{2\pi}{\tau} })^{\frac{26-d}{24}}~.
\label{FLongitudinalOpenChannel}
\end{equation}

We can see now that the open-string result (\ref{FLongitudinalOpenChannel}) is in perfect agreement with the closed-string amplitude (\ref{FLongitudinalClosedChannel}), up to numerical constants. Demanding that these agree as well, we can deduce with the aid of (\ref{mulongitudinal})  that
\begin{equation}
N_p=2^{\frac{4-d}{4}}(2\pi)^{\frac{d-2p-3}{2}}\frac{K^2}{\nu G_s}L_s^{\frac{d-2p-4}{2}}.
\label{Nplong}
\end{equation}

Now that we have fixed the normalization of the longitudinal boundary state (\ref{StateLong}), we are prepared to calculate the leading long-distance correction to the metric and Kalb-Ramond fields.

\subsection{One-point functions}
\label{expectlongsubsec}
For  a stack of coincident D$p$-branes located at $y^i=0$ ($i=p+1,\ldots,d-1$), the nonvanishing one-point functions that give the leading long-distance behaviour of the metric and Kalb-Ramond fields (associated with the tensorial massless $w=0$ states in the NR string spectrum) are
\begin{align}
\bra{0,w=0,k}\alpha_{1}^{i}\alpha_{1}^{j}D\ket{B^0_{\mbox{\tiny long}}}
&=-S^{ij}N_{p}
(2\pi)^{p+1}\delta^{(p+1)}(k_{\para})
\frac{1}{k_{\perp}^{2}
}~, 
\nonumber\\
\bra{0,w=0,k}\gamma_{1}\gtil_{1}D\ket{B^0_{\mbox{\tiny long}}}
&=\frac{2}{\mu}N_p
(2\pi)^{p+1}\delta^{(p+1)}(k_{\para})\frac{1}{k_\perp^2
}~, 
\\
\bra{0,w=0,k}\beta_{1}\btil_{1}D\ket{B^0_{\mbox{\tiny long}}}
&=\frac{\mu}{2}N_p
(2\pi)^{p+1}\delta^{(p+1)}(k_{\para})\frac{1}{k_\perp^2
}~.
\nonumber
\end{align}
Performing the Fourier transform to find the spatial dependence of the metric and Kalb-Ramond perturbations, we find
\begin{align}
\alpha_1^i\alpha_1^j ~~&\rightarrow  ~~-S^{ij}N_p\frac{1}{(d-3-p)\Omega_{d-2-p}\,r_{\perp}^{d-3-p}}~,
\nonumber\\
\gamma_1\gtil_1 ~~&\rightarrow ~~~~~~~\frac{2}{\mu}N_p\frac{1}{(d-3-p)\Omega_{d-2-p}\,r_{\perp}^{d-3-p}}~,
\label{1ptslongitudinal}\\
\beta\btil ~~&\rightarrow ~~~~~~~\frac{\mu}{2}N_p\frac{1}{(d-3-p)\Omega_{d-2-p}}\,r_{\perp}^{d-3-p}~,
\nonumber
\end{align}
where
\begin{equation}
\Omega_q\equiv\frac{2\pi^{\frac{q+1}{2}}}{\Gamma\left(\frac{q+1}{2}\right)}
\label{Omega}
\end{equation}
is the area of a $q$-dimensional unit sphere.

 For our intended comparison against the background of the next subsection, it is convenient to rewrite  the normalization constant (\ref{Nplong}) in the form
\begin{equation}
N_p=2^{\frac{2-d}{4}}\frac{R_D^{d-3-p}(d-3-p)\,\Omega_{d-2-p}}{\kappa}~,
\label{normalizationrewrite}
\end{equation}
so as to cancel out the factors $(d-3-p)\,\Omega_{d-2-p}$ in the denominator of (\ref{1ptslongitudinal}). 
For the reason to be explained in the next paragraph, in (\ref{normalizationrewrite}) we factor out the gravitational coupling constant $\kappa$, defined by the usual expression
\begin{equation}
2\kappa^2 \equiv  (2\pi)^{d-3}G_s^2 L_s^{d-2}~.
\label{kappa}
\end{equation}
The rewriting (\ref{normalizationrewrite}) thus amounts to defining the length scale $R_D$ through 
\begin{equation}
R_D^{d-3-p}\equiv L_s^{d-3-p}\frac{K^2}{\nu}
\frac{2^{d-4-p}\pi^{\frac{d-5-p}{2}}}{d-3-p}\Gamma\left(\frac{d-1-p}{2}\right)~.
\label{rd}
\end{equation}

It is important to observe now that the one-point functions (\ref{1ptslongitudinal}) refer to  the \emph{canonically-normalized} graviton field $h^{\mbox{\tiny can}}$, because the propagator was so normalized. To compare against the result of the next subsection, we need to rescale to the dimensionless graviton field $h$ as seen in the string action. This rescaling involves a factor of $\kappa$, which explains why we chose to separate out such factor in (\ref{normalizationrewrite}).
The precise normalization determined by the state-operator mapping is 
\begin{equation}
h_{ij}= 2\kappa h_{ij}^{\mbox{\tiny can}}~,
\qquad
h_{\gamma\gtil}=-\kappa h_{\beta\btil}^{\mbox{\tiny can}}~,
\qquad
h_{\beta\btil}=-\kappa h_{\gamma\gtil}^{\mbox{\tiny can}}~.
\end{equation}

Putting this all together, we finally predict that the long-distance massless background generated by the longitudinal D-brane stack is 
\begin{subequations}\label{hbordelong:joint}
\begin{align}
h_{\gamma\gtil}&=-\frac{\mu}{2^{\frac{2+d}{4}}}\frac{ R_D^{d-3-p}}{r_\perp^{d-3-p}}~,
\\
h_{\beta\btil}&=-\frac{2^{\frac{6-d}{4}}}{\mu}\frac{R_D^{d-3-p}}{ r_\perp^{d-3-p}}~, \\
h_{ij}&=-2^{\frac{6-d}{4}}S_{ij}\frac{R_D^{d-3-p}}{r_\perp^{d-3-p}}~.
\end{align}
\end{subequations}

\subsection{Background description}
\label{bkgdlongsubsec}

Next we will study the same longitudinal system in its alternative description as a black brane background. Here, the only excitations are the closed strings propagating through the curved geometry. The background presented in \cite{Gopakumar:2000na,Harmark:2000wv} as the supergravity dual of NCOS$_{p+1}$  was derived by taking the NR limit of a black brane in the parent string theory sourced by $K$ longitudinal D-branes with F1 charge density $\nu$. In the notation of \cite{Danielsson:2000mu}, it has the string-frame metric
\begin{align}
\frac{ds^2}{l_s^2} &= \frac{1}{L_s^2}H^{\frac{1}{2}}\frac{K^2}{G_s^2\nu^2}\frac{r^{7-p}}{R_D^{7-p}}\left( -dx_0^2+dx_1^2  \right) +\frac{1}{L_s^2}H^{-\frac{1}{2}}\left( dx_2^2+\dots +dx_p^2  \right) \nonumber \\  
&   \qquad+\frac{1}{L_s^2}H^{\frac{1}{2}}\left( dx_{p+1}^2+\dots + dx_9^2  \right)~,\label{fondolong}
\end{align}
Kalb-Ramond field
\begin{equation}
\frac{B_{01}}{l_s^2}=\frac{1}{G_s^2L_s^2}\frac{K^2}{\nu^2}\frac{r^{7-p}}{R_D^{7-p}}\label{fondoBlong}
\end{equation}
and dilaton field
\begin{equation}
g^2_{\mbox{\tiny eff}}\equiv g_s^2 e^{2\phi}=\frac{K^2}{\nu^2}H^{\frac{5-p}{2}}
\left(\frac{r}{R_D}\right)^{7-p}~,
\label{fondophilong}
\end{equation}
where
\begin{align}
H&=1+\left(\frac{R_D}{r}\right)^{7-p}~,
\nonumber\\
r^2&=x_{p+1}^2+\dots +x_{9}^2~,
\label{harmoniclong}\\
R_D^{7-p}&=\frac{1}{(7-p)}\frac{K^2}{\nu}\frac{(2\pi)^{7-p}L_s^{7-p}}{\Omega_{8-p}}~,
\nonumber
\end{align}
with the unit-sphere volume $\Omega_{8-p}$ defined by (\ref{Omega}). Of note here is the fact that the curvature radius $R_D$ included in (\ref{harmoniclong}) coincides with the $d=10$ version of the length scale defined in (\ref{rd}), associated with the metric sourced by the D-branes.

In the calculations we performed using the boundary state  formalism, the metric components are obtained as coefficients of vertex operators that are expected to be present directly in the deformed action. For this reason, the most direct way  to make the comparison with the lowest-order correction in the black brane description is to observe  the  form of the relativistic string action on this background,  far from the brane throat. By doing so, we expect to be able to  reproduce  the Gomis-Ooguri action   
(\ref{actiongo}) plus a small correction. Using this procedure we will recognize the graviton vertex operators in the correction term, and we will be able to read off the metric components from their coefficients.

Let us now run through this calculation in detail. Recall first that, in the limit that leads to NCOS theory on flat spacetime, the fields of the parent string theory are scaled with a parameter 
\begin{equation}
    \delta\equiv \frac{l_s^2}{L_s^2} \rightarrow 0
    \label{delta}
\end{equation} according to \cite{Danielsson:2000gi,Gomis:2000bd}
\begin{equation}
\frac{g_{ab}}{l_s^2}=\frac{1}{L_s^2}\delta^{-1}\eta_{ab}~, ~~~\frac{g_{ij}}{l_s^2}=\frac{1}{L_s^2}\delta_{ij}~, ~~~\frac{B_{01}}{l_s^2}=\frac{1}{L_s^2}(\delta^{-1}-\mu)~,
~~~g_s = \delta^{-1/2}G_s~,
\label{ncoslimit}
\end{equation}
with $\mu$ for the longitudinal D-brane case having the value (\ref{mulongitudinal}). 

Next, we make use of the fact that a relativistic string propagating on the non-trivial metric (\ref{fondolong}) and antisymmetric field  (\ref{fondoBlong}) is of course described by the familiar sigma-model action
\begin{equation}
S_{\sigma\mbox{\tiny -model}}=\frac{1}{2\pi l_s^2}\int d^2 z\left(  g_{ab}\partial X^a\partialbar X^b -B_{ab}\partial X^a\partialbar X^b +g_{ij}\partial X^i\partial X^j\right)~.
\label{sigmamodel}
\end{equation}
Here $l_s$ is the string length in the parent theory. 
Far from the brane throat, in the region where $r\gg R_D$, the field profiles (\ref{fondolong}) and (\ref{fondoBlong}) contain the small parameter
\begin{equation}
\delta^\prime \equiv \left(\frac{R_D}{r}\right)^{7-p}\ll 1~,
\end{equation}
which, following \cite{Danielsson:2000mu}, we will want to interpret as playing a role analogous to $\delta$ in (\ref{ncoslimit}). In other words, we anticipate that, in the supergravity background (\ref{fondolong})-(\ref{fondoBlong}), the effect of moving out to $r\to\infty$ is what implements the NR limit.

We are thus interested in examining the manner in which (\ref{sigmamodel}) simplifies in this long-distance limit. 
We expect to have
\begin{equation}
S_{\sigma\mbox{\tiny -model}}= S_{\mbox{\tiny GO}}+O(\delta)~. \label{dSlong}
\end{equation}
The leading-order terms were already shown to arise in \cite{Danielsson:2000mu}; 
here we are interested in working out the $O(\delta)$ terms. 
To have a match in this sense between (\ref{sigmamodel}) and the scaling (\ref{ncoslimit}) that leads to (\ref{actiongo}), we must identify
\begin{equation}
L_s^2\frac{B_{01}}{l_s^2}=2\mu\left(\frac{r}{R_D}\right)^{7-p}\equiv \delta^{-1}-\mu +O(\delta)~,\label{condB1}
\end{equation}
\begin{equation}
L_s^2\frac{g_{ab}}{l_s^2}=2\mu H^{\frac{1}{2}}\left( \frac{r}{R_D}\right)^{7-p}\eta_{ab}\equiv\left(\delta^{-1}+O(\delta^0)\right)\eta_{ab}~, \label{condB2}
\end{equation}
and 
\begin{equation}
g_{\mbox{\tiny eff}}^2=2\mu G^2_s H^{\frac{5}{2}}\left( \frac{r}{R_D}\right)^{7-p}=\frac{G_s^2}{\delta}+O(\delta^0)~.
\label{condB22}
\end{equation}
Notice that it is not obvious  a priori that these three conditions can be satisfied simultaneously. In more detail, they read
\begin{align}
\frac{2\mu}{\delta^\prime}&=\frac{1}{\delta}-\mu+a\delta~, \label{condB3}
\\
\left( 1+\frac{1}{2}\delta^\prime-\frac{1}{8}\delta^{\prime 2}\right)\frac{2\mu}{\delta^\prime}&=\frac{1} {\delta}+b+c\delta~,\label{condB4}
\\
\frac{2\mu}{\delta'}+(5-p)\mu +\frac{1}{4}(5-p)(4-p)\mu\delta'&=\frac{1}{\delta}+e+f\delta~,
\label{condB5}
\end{align}
where $a,b,c,e,f$ are parameters to be determined.
Demanding compatibility between
(\ref{condB3})-(\ref{condB5}), we learn that 
\begin{equation}
b=0~, \quad 
c=a-\frac{1}{2}\mu^2~, \quad 
e=(4-p)\mu~, \quad
f=a+\frac{(5-p)(4-p)}{2}\mu^2~.
\label{a-b}
\end{equation}
At this stage, $a$ is still undetermined. In other words, the asymptotic behavior required in  (\ref{condB1})-(\ref{condB22}) for the string to be automatically and correctly subjected to the NR limit (\ref{ncoslimit}) as $r\to\infty$ holds irrespective of the value of $a$. 

With this information, we get the $O(\delta)$ correction to $\frac{g_{ij}}{l_s^2}$ univocally as
\begin{equation}
\frac{g_{ij}}{l_s^2}\simeq \frac{1}{L_s^2}\left(1-S_{ij}\frac{1}{2}\delta^\prime\right)\simeq \frac{1}{L_s^2}\left(1- S_{ij}\mu\delta\right)~.
\label{correcgijlong}
\end{equation}
where $S_{ij}$ is given by (\ref{Sijlong}). Using (\ref{correcgijlong}), the transverse part of the action reads
\begin{equation}
S_{\mbox{\tiny trans}} \simeq \int\frac{d^2z}{2\pi L_s^2}(\eta_{ij}-S_{ij}\mu\delta )\partial X^i\bar\partial X^j =\int\frac{d^2z}{2\pi L_s^2}\partial X^i\bar\partial X^i +\delta\int\frac{d^2z}{2\pi L_s^2}(-\mu S_{ij})\partial X^i\bar\partial X^j~.
\label{stranslong}
\end{equation}
When  $\delta\rightarrow 0$, this action takes  the form (\ref{dSlong}), and we can easily read off the corresponding correction to the metric. 

The longitudinal part of the action reads
\begin{align}
S_{\mbox{\tiny long}}&=\int\frac{d^2z}{2\pi L_s^2}\left[\frac{g_{ab}}{l_s^2}\partial X^a\partialbar X^b -\frac{B_{ab}}{l_s^2}\partial X^a\partial X^b]\right] \nonumber\\
&= \int\frac{d^2z}{2\pi L_s^2}\left[ \left(1+\frac{1}{2}\delta^\prime -\frac{1}{8}\delta^{\prime 2}\right)\frac{2\mu}{\delta^\prime}\eta_{ab}\partial X^a \partialbar X^b -\frac{2\mu}{\delta^\prime}\epsilon_{ab}\partial X^a \partialbar X^b\right]~. 
\end{align}
Expressing it in terms  of $\gamma$  and $\tilde\gamma$, we have
\begin{align}
S_{\mbox{\tiny long}} &=\int\frac{d^2z}{2\pi }\Big[\frac{1}{2\delta}(\partial \gamma\partialbar\gtil +\partial\gtil\partialbar\gamma)-\frac{1}{2}\left(\frac{1}{\delta}-\mu\right)(\partial\gamma\partialbar\gtil -\partial\gtil\partialbar\gamma)\nonumber \\
& \qquad\qquad+\frac{c\delta}{2}(\partial\gamma\partialbar\gtil +\partial\gtil\partialbar\gamma)-\frac{a\delta}{2}(\partial\gamma \partialbar\gtil -\partial\gtil\partialbar\gamma)\Big]~.
\end{align}
Introducing as in \cite{Gomis:2000bd} Lagrange multipliers $\beta$, $\tilde\beta$, we can rewrite the longitudinal action as
\begin{align}
S_{\mbox{\tiny long}} &=\int\frac{d^2z}{2\pi }\Big[\frac{\mu}{2}\partial\gamma\bar\partial \tilde\gamma +\beta\bar\partial\gamma +\tilde\beta\partial\tilde\gamma -\frac{1}{\left(\frac{1}{\delta}-\frac{\mu}{2}\beta\tilde\beta\right)}\nonumber \\
& \qquad\qquad+\frac{c\delta}{2}(\partial\gamma\partialbar\gtil +\partial\gtil\partialbar\gamma)-\frac{a\delta}{2}(\partial\gamma \partialbar\gtil -\partial\gtil\partialbar\gamma)\Big]~.
\end{align}

So, finally when $\delta\rightarrow 0$ we get (to first order)
\begin{align}
S_{\mbox{\tiny long}} &=\int\frac{d^2z}{2\pi}\Big[ \beta\partialbar\gamma +\btil\partial\gtil +\frac{\mu}{2}\partial\gamma\partialbar\gtil  \nonumber \\
&  \qquad\qquad-\frac{\delta}{2}(a-c)\partial\gamma\partialbar\gtil +\frac{\delta}{2}(a+c)\partial\gtil\partialbar\gamma -\delta\beta\btil\Big]~.~~~~~~~~~~
\end{align}
This expression has the desired form (\ref{dSlong}). The value of $a-c$ was fixed in (\ref{a-b}), but that of $a+c$ has remained until now undetermined. The appropriate choice for comparing with the boundary state result,
where $\partial\tilde\gamma=\bar\partial\gamma=0$ due to the equations of motion, is 
\begin{equation}
a+c= 0 \quad \longleftrightarrow \quad a=\frac{1}{4}\mu^2~.
\end{equation}
Together with (\ref{a-b}), this entails that 
\begin{equation}
\delta=\frac{\delta'}{2\mu}-\frac{\delta'^2}{4\mu}+\frac{5\delta'^3}{32\mu}+O(\delta'^4)~.
\label{deltadeltaprime}
\end{equation} 

With this information, we finally obtain
\begin{align}
S_{\mbox{\tiny long}} &=\int\frac{d^2z}{2\pi}\Big[ \beta\partialbar\gamma +\btil\partial\gtil +\frac{\mu}{2}\partial\gamma\partialbar\gtil -\frac{\delta}{4}\mu^2\partial\gamma\partialbar\gtil  -\delta\,\beta\btil\Big]~.~~~~~~~~~~
\end{align}
From this result, together with (\ref{stranslong}) and the leading term in (\ref{deltadeltaprime}), we find that the coefficients of
\begin{equation}
\partial\gamma\partialbar\gtil  ,~~~~~~~ \beta\btil,~~~~~~~\partial X^i\partialbar X^i~,
\end{equation}
to first order in $(R_D/r)^{7-p}$, are given respectively by
\begin{subequations}\label{hfondolong:joint}
\begin{align}
h_{\gamma\tilde\gamma}&= -\frac{\mu}{8}\left( \frac{R_D}{r} \right)^{7-p}~,
\label{coeflong1}
\\
h_{\beta\tilde\beta}&=-\frac{1}{2\mu}\left( \frac{R_D}{r} \right)^{7-p}~ , 
\label{coeflong2}\\
h_{ij}&=-\frac{1}{2}S_{ij}\left( \frac{R_D}{r}\right)^{7-p}~.
\label{coeflong3}
\end{align}
\end{subequations}
If we substitute $d=10$ in 
the boundary state result (\ref{hbordelong:joint}), we obtain perfect agreement with (\ref{hfondolong:joint}).

\section{Transverse D-branes}
\label{transversesec}
Next we will run through the same calculation for a stack of transverse D-branes. Along the way, we will highlight the differences with respect to the previous section that are significant, but we will be brief in the steps that are identical. 

\subsection{Boundary conditions}
\label{bctranssubsec}

By definition, a transverse D-brane is localized along $X^1=L_s(\gamma+\tilde\gamma)/2$. Through (\ref{fronteraWSTabierta}), the variational principle yields the    boundary conditions \cite{Danielsson:2000mu}
\begin{equation}
(\gamma+\tilde\gamma)|_{z=\bar z}=\text{constant}~,\quad (\beta+\tilde\beta)|_{z=\bar z}=0~~.\label{condfronttrans}
\end{equation}
These go together with the standard boundary conditions for the $X^i$ embedding fields:  Neumann along the $p$ spatial directions parallel to the D-brane stack ($i=2,\ldots,p+1$), and Dirichlet along the directions which, together with $x^1$, are orthogonal to the brane ($i=p+2,\ldots,d-1$). Just like on-shell closed strings, in NR string theory such open strings must have positive winding \cite{Danielsson:2000gi,Danielsson:2000mu}.

To write out the first condition in (\ref{condfronttrans}) more precisely, consider an open string stretched between two transverse D-branes. There are as usual two solutions that differ in orientation. For an  open string with winding  number $w$ that begins  on a D-brane located at $x^1=0$ and ends on a D-brane  at  $x^1=y^1$, we have
\begin{equation}
\frac{L_s}{2}(\gamma+\gtil)|_{z=\bar{z}}=
\left\{
\begin{array}{ll}
0  &  z>0 \\
y^1+2\pi Rw & z<0
\end{array} 
\right.
,\quad (\beta+\btil)|_{z=\bar{z}}=0 ~.
\label{bctransdetailed}
\end{equation}
If we define the fractional winding  $w^\prime=w+\frac{y^1}{2\pi R}$ as in \cite{Danielsson:2000mu}, then
the solution for the longitudinal embedding fields that meets conditions (\ref{bctransdetailed}) reads  
\begin{align}
\gamma(z)&=-i\frac{2w^\prime R}{L_s}\log z+\sum_n\gamma_n z^{-n}~, 
\nonumber\\
\gtil(\bar{z})&=~i\frac{2w^\prime R}{L_s}\log\bar{z}-\sum_n\gamma_n\bar{z}^{-n}~,\\
\beta(z)&=\sum_n\beta_n z^{-n-1}~~,
\nonumber\\
\btil(\hat{z})&=-\sum_n\beta_n \bar{z}^{-n-1}~.
\nonumber
\end{align}
These are complemented by the standard mode expansions for the $X^i$. 

The spectrum is 
\cite{Danielsson:2000mu}
\begin{equation}
 p_0=\mu\frac{w^\prime R}{L_s^2}+\frac{L_s^2 p_\perp^2}{2w' R}+\frac{N_{\mbox{\tiny long}}+N_{\mbox{\tiny trans}}}{2w^\prime R}~.
\end{equation}
Just as for closed strings \cite{Danielsson:2000gi}, for open strings on transverse D-branes the $\mu$ parameter in the action is totally arbitrary \cite{Danielsson:2000gi,Danielsson:2000mu}, and we can choose in particular $\mu=0$. Then
\begin{align}
L_0^{\mbox{\tiny long}}&=-2p_0 w^\prime R+N_{\mbox{\tiny long}} ~,\label{L0transparWST}\\
L_0^{\mbox{\tiny trans}}&=L_s^2p_ip^i+\frac{y^2}{(2\pi )^2L_s^2}+N_{\mbox{\tiny trans}} ~.\label{L0transperpWST}
\end{align}

For the remaining sector, where the string has the initial endpoint on the D-brane at  $x^1=y^1$ and the final endpoint attached to the D-brane located at $x^1=0$, we  have
\begin{equation}
\frac{L_s}{2}(\gamma+\gtil)|_{z=\bar{z}}=
\left\{
\begin{array}{ll}
y^1  &  z>0 \\
2\pi w R & z<0
\end{array}
\right.  ~.
\label{pormientras}
\end{equation}
It is convenient to report this a bit differently, by making use of  the periodicity condition $x^1\simeq x^1+2\pi$ to shift the endpoint locations backwards, to $y^1-2\pi wR$ and $0$. If we now define the fractional winding as  $w^{\prime\prime}=w-\frac{y^1}{2\pi R}$ (such that the positive winding condition $w>0$ still implies $w^{\prime\prime}>0$), (\ref{pormientras}) becomes
\begin{equation}
\frac{L_s}{2}(\gamma+\gtil)|_{z=\bar{z}}=
\left\{
\begin{array}{ll}
-2\pi w^{\prime\prime} R  &  z>0 \\
0 & z<0
\end{array}
\right.  ~,
\end{equation}
leading to the mode expansions
\begin{equation}
\begin{split}
\gamma(z)&=-\frac{2\pi w^{\prime\prime} R}{L_s}-i\frac{2w^{\prime\prime} R}{L_s}\log z+\sum_n\gamma z^{-n}~, \\
\gtil(\bar{z})&=-\frac{2\pi w^{\prime\prime} R}{L_s}+i\frac{2 w^{\prime\prime} R}{L_s}\log\bar{z}-\sum_n\gamma \bar z^{-n}~.
\end{split}
\end{equation}

In this sector (\ref{L0transparWST})
changes to
\begin{equation}
L_0^{\mbox{\tiny long}}=-2p_0 w^{\prime\prime} R +N_{\mbox{\tiny long}} ~,
\end{equation}
and (\ref{L0transperpWST}) is unchanged.

\subsection{Boundary state}
\label{bdrystatetranssubsec}
In this section we calculate the boundary state for a stack of coincident transverse D-branes,  analogously to the procedure used in Section~\ref{bdrystatelongsubsec}. An important difference is that, now that the D-branes do not span the $x^1$ direction, closed string winding is conserved all by itself: even if the string breaks open, there is no possibility for it to unwind. This implies that a closed string emitted by a transverse D-branes must necessarily have zero winding. This is T-dual to the perhaps more familiar statement that, on account of translational invariance, a closed string emitted by a D-brane must have zero momentum along all directions spanned by the brane. This requirement was visible in the previous section, and will naturally also be relevant for the analysis that follows.  

Letting $x^i=y^i$ with $i=1,p+2,\ldots,d-1$ denote the location of the stack, and focusing on open strings with $w=0$ for the reason explained in the previous paragraph,\footnote{Just like in the previous section, the positive-winding condition is unenforced at this level where we are again discussing an off-shell process.} regardless of orientation we are led to consider the longitudinal boundary conditions
\begin{equation}
\begin{split}
\frac{L_s}{2}(\gamma +\gtil)\Big{|}_{z=\zbar} &= y^1~, \\
(\beta+\btil)\Big{|}_{z=\zbar} &= 0~,
\end{split}
\label{bctrans}
\end{equation}
which will be of course supplemented below by the usual boundary conditions for the transverse embedding fields.

Performing  as before a conformal transformation that exchanges $\sigma\leftrightarrow\tau$, the boundary conditions (\ref{bctrans}) are translated into the following initial conditions for the boundary state:
\begin{equation}
\begin{split}
\frac{L_s}{2}\left( \gamma +\gtil\right)\Big{|}_{z={\zbar}^{-1}}\ket{B_{\mbox{\tiny trans}}}&=y^1\ket{B_{\mbox{\tiny trans}}}~,\\
\left( z\beta(z)-\zbar\btil(\zbar)\right)\Big{|}_{z={\zbar}^{-1}}\ket{B_{\mbox{\tiny trans}}}&=0~.
\end{split}
\end{equation}
Expressing these conditions in terms of the  mode expansion for the closed string, they read 
\begin{equation}
\begin{split}
w\ket{B_{\mbox{\tiny trans}}} &=0~, 
\\
\frac{L_s}{2}(\gamma_0 +\gtil_0)\ket{B_{\mbox{\tiny trans}}} &=y^1\ket{B_{\mbox{\tiny trans}}} \quad 
\\
(\gamma_n +\gtil_{-n})\ket{B_{\mbox{\tiny trans}}} &= 0\qquad \forall\,n\neq 0~,  
\label{bdrystatetransconditions}\\
(\beta_n -\btil_{-n})\ket{B_{\mbox{\tiny trans}}} &= 0 \qquad \forall\,n\neq 0~, \\
(\beta_0-\btil_0)\ket{B_{\mbox{\tiny trans}}} &= 0~.
\end{split}
\end{equation}
The first condition expresses the fact that the winding must vanish, as we had argued before. Taking that into account, the final condition can be reexpressed as 
\begin{equation}
p_0\ket{B_{\mbox{\tiny trans}}}=0~,
\end{equation}
which is the expected statement that, due to invariance under time translations, the D-brane can only emit closed strings with zero energy. 

As a consequence of having compactified  $x^1$ on a circle, we can either restrict the eigenvalues of $\hat x^1$ to lie in the $(0,2\pi R)$ range, or go to the covering space of the circle and place copies of the D-brane stack at all points $x^1= y^1+2\pi mR$ with integer $m$. In the latter perspective, the second condition in (\ref{bdrystatetransconditions}) is then written out as
\begin{equation}
\begin{split}
    \ket{B_{\mbox{\tiny trans}}}&=\sum_{m=-\infty}^{\infty} \ket{B^{m}_{\mbox{\tiny trans}}}
    \\
 \hat{x}^1\ket{B^{m}_{\mbox{\tiny trans}}}&=(y^1+2\pi mR)\ket{B^{m}_{\mbox{\tiny trans}}}
 \label{Bcondperiod} 
 \end{split}
\end{equation}

For the $X^i$ coordinates, the conditions are as usual
\begin{equation}
\begin{split}
(z\partial +\bar z\bar\partial)X^i|_{z=\bar z^{-1}}\ket{B_{\mbox{\tiny trans}}} &= 0\quad \quad i= 2,\dots, p+1 ~, \\
X^i|_{z=\bar z^{-1}}\ket{B_{\mbox{\tiny trans}}} &= y^i\ket{B_{\mbox{\tiny trans}}}\quad ;\quad i= p+2,\dots ,d-1~.
\end{split}
\end{equation}
Or, in  terms of mode expansions, 
\begin{equation}
\begin{split}
(\alpha^i_n +S^i_j\tilde{\alpha}^j_{-n})\ket{B_{\mbox{\tiny trans}}}&=0\quad ;\quad n\neq 0~,\\
\hat{p}^{\alpha}_{\para}\ket{B_{\mbox{\tiny trans}}}&=0 ~,\\
(\hat{x}^i_{\perp}-y^i)\ket{B_{\mbox{\tiny trans}}}&=0~,
\end{split}
\end{equation}
where we have defined the $(d-2)\times(d-2)$ matrix
\begin{equation}
S^{ij}\equiv\left\{ \begin{array}{ll}
~~\delta^{ij}  &  \text{if } \ \ i,j =2,\dots , p+1\\
-\delta^{ij} & \text{if } \ \ i,j =p+2,\dots ,d \\
\end{array}\right.  ~.\label{Sijtrans}
\end{equation}

Using all of the above initial conditions, the boundary state for a transverse D-brane is determined to be
\begin{align}
\ket{B_{\mbox{\tiny trans}}}&=N_p\sum_m\delta(\hat{x}^1-2\pi m R)\,\delta^{d-p-2}(\hat{x}^i-y^i)\nonumber \\
&\hspace{1.3cm}\prod_{n-1}^\infty e^{-(\frac{1}{n}\alpha_{-n}S\cdot\tilde{\alpha}_{-n}+\beta_{-n}\gtil_{-n}+\btil_{-n}\gamma_{-n})}\ket{0,w=0,p=0}~.~~~\label{BtransWST1}
\end{align}
Proceeding in the same manner as we did for the longitudinal D-brane, we can determine the normalization of the boundary state comparing the interaction of two parallel D-branes in the
open and closed string channels. In the end, we obtain
\begin{equation}
N_p^2=K^2L_s^{d-2p-4}2^{\frac{4-d}{2}}(2\pi)^{d-2p-3}~.
\label{Nptrans}
\end{equation}

\subsection{One-point functions}
\label{expectranssubsec}
Projecting $D|B_{\mbox{\tiny trans}}\rangle$ onto the massless closed string states, the non-vanishing results are the following:
\begin{align}
\bra{0,w=0,k_1=\frac{m}{R}, k_\alpha,k_\perp}\gamma_1 \btil_1D\ket{B_{\mbox{\tiny trans}}}
&=\frac{N_p L_s^2}{8\pi}\int\limits_{|z|\leq 1}\frac{d^2z}{|z|^2}|z|^{\frac{L_s^2}{2}k_\perp^2}(2\pi)^{p+1}\delta^{(p+1)}(k_\alpha) ~,
\nonumber\\
&=-N_p (2\pi)^{p+1}\delta^{(p+1)}(k_\alpha)\frac{1}{k_\perp^2}~,
\label{1pttrans}\\
\bra{0,w=0,k_1=\frac{m}{R}, k_\alpha,k_\perp}\gtil_1\beta_1 D\ket{B_{\mbox{\tiny trans}}}
&=-N_p (2\pi)^{p+1}\delta^{(p+1)}(k_\alpha)\frac{1}{k_\perp^2}~,
\nonumber\\
\bra{0,w=0,k_1=\frac{m}{R}, k_\alpha,k_\perp}\alpha_1^i\tilde{\alpha}_1^jD\ket{B_0}&=-N_p (2\pi)^{p+1}\delta^{(p+1)}(k_\alpha)\frac{S^{ij}}{k_\perp^2}~.
\nonumber
\end{align}
Performing a Fourier transform to pass to the coordinate representation, we obtain the metric components 
\begin{subequations}
\label{htrans}
\begin{align}
h_{\gamma\tilde{\beta}}&=h_{\tilde{\gamma}\beta}=-\kappa h_{\tilde{\gamma}\beta}^{\mbox{\tiny can}}\nonumber\\
&=
\frac{2^{\frac{2-d}{4}}(2\pi)^{d-3-p}KG_s}{(d-p-4)\Omega_{d-3-p}}\frac{L_s^{d-p-3}}{r_\perp^{d-p-4}}
\sum_n\delta(x^1-2\pi nR)~,
\label{Bmetricabgtrans}
\\  
h_{ij}&=2\kappa h_{ij}^{\mbox{\tiny can}}\nonumber\\
&=-
\frac{2^{\frac{6-d}{4}}(2\pi)^{d-3-p}KG_s}{(d-p-4)\Omega_{d-3-p}}\frac{S_{ij}L_s^{d-p-3}}{r_\perp^{d-p-4}}
\sum_n\delta(x^1-2\pi nR)~,
\label{Bmetricaxtrans} 
\end{align}
\end{subequations}
where we have substituted  (\ref{Nptrans}) and (\ref{kappa}).
Note that in (\ref{1pttrans}) and (\ref{htrans}) the  $\perp$ subindex refers only to those components perpendicular to the D-branes that are non-longitudinal (i.e., $i=p+2,\ldots,d-1$).

\subsection{Background description}
\label{bkgdtranssubsec}
The background corresponding to transverse D-branes in NR string theory has not been presented elsewhere, so we will derive it here. To this aim, we start with the standard black $p$-brane solution of relativistic string theory \cite{Horowitz:1991cd},  
\begin{align}
    ds^2&= H^{-1/2}\left(-d\tilde{x}_0^2+d\tilde{x}_2^2+\ldots+d\tilde{x}_{p+1}^2\right)
    +H^{1/2}\left(d\tilde{x}_1^2+d\tilde{x}_{p+2}^2+\ldots+d\tilde{x}_9^2\right)~,
\nonumber\\
g_{\mbox{\tiny eff}}^2 &= g_s^2 e^{2\phi}=g_s^2 H^{\frac{3-p}{2}}~,
\label{blackpbrane}\\
H&=1+ \frac{c_p K g_s l_s^{7-p}}{\left(x_1^2+\tilde{r}_{\perp}^2\right)^{(7-p)/2}}~,
\nonumber\\
c_p&\equiv (4\pi)^{\frac{5-p}{2}}\Gamma\left(\frac{7-p}{2}\right)~,
\nonumber
\end{align}
where $g_s$ and $l_s$ are the string coupling and string length, and we have labeled the coordinates with tildes in anticipation of the rescaling we will perform when taking the NR limit. To prepare further for this limit, without affecting (\ref{blackpbrane})
we can turn on a constant Kalb-Ramond field
\begin{equation}
    B_{01}=1-\mu\frac{l_s^2}{L_s^2}~. 
\label{antisymmetric}
\end{equation}
Finally, to compactify the solution (\ref{blackpbrane}) along $x^1$, we need to generalize the harmonic function $H$ to a sum of terms that incorporate mirror images of the D-brane stack at $x^1=2\pi m R$ for all integer $m$,
\begin{equation}
    H=1+ \sum_{m=-\infty}^{\infty}\frac{c_p K g_s l_s^{7-p}}{\left[\left(x_1-2\pi m R\right)^2+\tilde{r}_{\perp}^2\right]^{(7-p)/2}}~.
\end{equation}

Now we take the NR limit (\ref{delta}), implementing (\ref{ncoslimit}) through the scalings
\begin{equation}
\tilde{x}^a=x^a~,\qquad
\tilde{x}^i=\sqrt{\delta}\,x^i~,\qquad
g_s=\frac{G_s}{\sqrt{\delta}}~.
\end{equation}
This converts (\ref{blackpbrane}) and (\ref{antisymmetric}) into\footnote{We note that the sum over $m\in\bZ$ inside the harmonic function $H$ contains a delta function that should not be confused with the small parameter $\delta$.}
\begin{equation}
\begin{split}
\frac{ds^2}{l_s^2}&= \frac{1}{L_s^2}\biggl[ -\frac{1}{\delta}H^{-\frac{1}{2}}dx_0^2+\frac{1}{\delta}H^{\frac{1}{2}}dx_1^2 +H^{-\frac{1}{2}}(dx^2_2+\dots +dx_{p+1}^2) 
\\
&\quad\quad\quad\quad\quad
+H^{\frac{1}{2}}(dx^2_{p+2}+\dots +dx^2_9)\biggr]~, 
\label{fondotrans}
\\
\frac{B_{01}}{l_s^2} &=\frac{1}{L_s^2}\left(\frac{1}{\delta}-\mu\right)~,
\\
g_{\mbox{\tiny eff}}^2 &= \frac{1}{\delta}G_s^2 H^{\frac{3-p}{2}}~,
\\
H &= 1+\left(\frac{R_D^{7-p}}{r_\perp^{6-p}}\right)\sum_{m=-\infty}^{\infty}
\delta\!\left(x^1-2\pi mR\right)~,
\\
R_D^{7-p}&\equiv 2^{5-p}\pi^{\frac{6-p}{2}}\,\Gamma\!\left(\frac{6-p}{2}\right)K G_s L_s^{7-p}~.
\end{split}
\end{equation}
Notice that the resulting structure here is very different from the one in the longitudinal black brane (\ref{fondolong}). The main distinction is the fact that, even after the limit, the parameter $\delta$ is still present in the right-hand side of the first three equations in (\ref{fondotrans}). This implies that the string worldsheet action on this background will only be non-singular after we give it the Lagrange-multiplier treatment of \cite{Gomis:2000bd}. For our purposes it will suffice to address this here in the region far from the brane throat. The rewriting of the complete background will be discussed elsewhere \cite{Avila:2023}. 

Another notable feature of (\ref{fondotrans}) is the presence of the delta function, which informs us that the NR limit has the effect of completely localizing the metric profile
along the longitudinal direction $x^1$, while keeping the profile nontrivial along the rest of the directions  perpendicular  to the brane. At any point that is even slightly off the source along $x^1$, the harmonic function is identically equal to one, and so (\ref{fondotrans}) is precisely the flat background for NR string theory studied in the original works \cite{Danielsson:2000gi,Gomis:2000bd}. On the other hand, the  background is nontrivial within the 9-dimensional plane at $x^1=0$. 

Far from the brane throat along the orthogonal non-longitudinal directions, we have in the harmonic function $H$ the small quantity
\begin{equation}
    \delta^\prime \equiv \left(\frac{R_D}{r_\perp}\right)^{7-p}~\sum_{m=-\infty}^{\infty}
    \delta\!\left(\frac{x^1-2\pi mR}{r_\perp}\right)~,
\label{delprimtrans}
\end{equation}
in terms of which we can expand
\begin{equation}
    H^{\pm\frac{1}{2}}= (1+\delta^\prime)^{\pm\frac{1}{2}}\simeq 1\pm\frac{1}{2}\delta^\prime~.
    \nonumber
\end{equation}

We emphasize that, unlike what happened in Section~\ref{bkgdlongsubsec}, the parameter $\delta$ for the transverse black brane is independent of $\delta^\prime$ defined in (\ref{delprimtrans}). As  we will see, in the limit far from the brane throat, $\delta$ does not appear in the expressions, so in the end, everything is given in terms of  $\delta^\prime$.

The action that results after inserting the metric and  Kalb-Ramond fields (\ref{fondotrans}) into the $\sigma$-model action is
\begin{equation}
\begin{split}
S_{\sigma\mbox{\tiny -model}}&=\frac{1}{2\pi L_s^2}\int d^2 z\Big[\delta^{-1}(-H^{-\frac{1}{2}}\partial X^0\partialbar X^0 +H^{\frac{1}{2}}\partial X^1\partialbar X^1)
\\
&\hspace{2.5cm}-(\delta^{-1}-\mu)(\partial X^0\partialbar X^0 -\partial X^1\partialbar X^0) 
\\
&\hspace{2.5cm}+H^{-\frac{1}{2}}(\partial X^2\bar\partial X^2+\cdots +\partial X^{p+1}\bar\partial X^{p+2})
\\
&\hspace{2.5cm}+H^{\frac{1}{2}}(\partial X^{p+2}\bar\partial X^{p+2}+\cdots +\partial X^{d-1}\bar\partial X^{d-1})\Big]~.
\end{split}
\end{equation}
Far from the throat, that is, when $\delta^\prime\rightarrow 0$,
\begin{align}
S_{\sigma\mbox{\tiny -model}}&\simeq \frac{1}{2\pi L_s^2}\int d^2 z\Big[\frac{\mu}{2}\partial \gamma\partialbar \gtil +\frac{1}{2}(2\delta^{-1}-\mu)\partial \gtil\partialbar\gamma +\frac{\delta^\prime}{4}\delta^{-1}(\partial\gamma\partialbar\gamma +\partial\gtil\partialbar\gtil)
\nonumber \\
&~~~~~~~~~~~~~~\qquad+\left(1-\frac{1}{2}\delta^\prime\right)(\partial X^2\bar\partial X^2+\cdots +\partial X^{p+1}\bar\partial X^{p+1}) \\
&~~~~~~~~~~~~~~\qquad+\left(1+\frac{1}{2}\delta^\prime\right)(\partial X^{p+2}\bar\partial X^{p+2}+\cdots +\partial X^{d-1}\bar\partial X^{d-1})\Big]~.
\nonumber
\end{align}
We observe here that the $X^i$ part of the action takes the desired form (Gomis-Ooguri + small correction of order $\delta^\prime$ ). As for the longitudinal part, since $\delta $ is the parameter involved in the definition of NR string theory, we must consider the limit $\delta\rightarrow 0$. To achieve this, we introduce Lagrange multipliers  $\beta,~\tilde\beta$ \cite{Gomis:2000bd}. The equivalent action in terms of these auxiliary variables (remaining consistently to linear order in $\delta'$) is
\begin{equation}
S_{\para} =\int\frac{d^2z}{2\pi}\Big[ \frac{\lambda}{2}\partial\gamma\partialbar\gtil +\btil\partialbar\gtil +\btil\partial\gtil -\frac{1}{\delta^{-1}-\frac{\lambda}{2}}\beta\btil +\frac{\delta^\prime\delta^{-1}}{4\left(\delta^{-1}-\frac{\lambda}{2}\right)}(\beta\partialbar\gtil +\btil\partial\gamma)\Big]~.
\end{equation}
We appreciate here that the action in terms  of $\beta$, $\tilde\beta$ does not become infinite when $\delta\rightarrow 0$.  Taking the limit, we conclude that the large-distance action is
\begin{align}
S&=S_{\mbox{\tiny GO}}+\frac{1}{2\pi L_s^2}\int d^2 z\left(\frac{\delta^\prime}{4}\beta\partialbar\gtil +\frac{\delta^\prime}{4}\btil\partial\gamma\right)
\nonumber\\
&~~~~~~+\frac{1}{2\pi L_s^2}\int d^2 z\left(-\frac{1}{2}\delta^\prime\right)(\partial X^2\bar\partial X^2+\cdots +\partial X^{p+1}\bar\partial X^{p+1})
\label{godeltaprime}\\
&~~~~~~~~~~~~~~+\frac{1}{2\pi L_s^2}\int d^2z\left(\frac{1}{2}\delta^\prime\right)(\partial X^{p+2}\bar\partial X^{p+2}+\cdots +\partial X^{d-1}\bar\partial X^{d-1})\Big]~.
\nonumber
\end{align}
As anticipated, this is the Gomis-Ooguri action plus a correction of order $\delta^\prime\ll 1$.

{}From (\ref{godeltaprime}) we can read off the coefficients that give the correction to the metric,
\begin{subequations}
\begin{align}
\left.
\begin{array}{l}
\beta\partialbar\gtil  \\
\btil\partial\gamma
\end{array}
\right\}~~
\rightarrow ~~h_{\gamma\tilde\beta}=h_{\tilde\gamma\beta}&=
\frac{1}{4}\delta'
\nonumber\\
~~~~&=\frac{(2\pi)^{7-p}KG_s}{4(6-p)\Omega_{7-p}}\frac{L_s^{7-p}}{ r_\perp^{6-p}}\sum_{m=-\infty}^{\infty}\delta(x^1-2\pi mR)~,
 \label{metricabgtrans}
 \end{align}
 \begin{equation}
 \partial X^i\partialbar X^j ~~~~\rightarrow ~~h_{ij}= -\frac{(2\pi)^{7-p}KG_s}{2(6-p)\Omega_{7-p}}\frac{S_{ij}L_s^{7-p}}{ r_\perp^{6-p}}\sum_m\delta(x^1-2\pi mR)~,
 \label{metricaxtrans}
 \end{equation}
 \end{subequations}
where $S^{ij}$ is the matrix defined in (\ref{Sijtrans}).

Now that we have obtained in (\ref{metricabgtrans}) and (\ref{metricaxtrans}) the lowest-order correction to the metric  from the perspective of the gravity background, we can compare them  with  our results in Section~\ref{bdrystatetranssubsec}, where we determined
the metric sourced by a boundary state that encodes a stack of $K$ transverse D-branes.  We  observe that, upon substituting $d=10$ in (\ref{Bmetricabgtrans}), (\ref{Bmetricaxtrans}), the boundary state metric and the black brane metric indeed match. We have thus verified the equivalence of the two descriptions.

\section*{Acknowledgements}
We are grateful to Daniel \'Avila for valuable conversations and helpful comments on the first version of this paper. 
Our work was partially supported by Mexico's National Council of Science and Technology (CONACyT) grant A1-S-22886 and DGAPA-UNAM grant IN116823.

\bibliography{DbraneNRgeometry}

\bibliographystyle{./utphys}

\end{document}